\documentstyle{l-aa}
\font\sixrm=cmr6
\newcommand{\magpt}[2]{\mbox{$\rm #1\hspace{-0.25em}\stackrel{m}{.}
   \hspace{-1.0mm}#2$}}               
\newcommand\ion[2]{\hbox{#1\,{\sixrm #2}}}
\newcommand\ebv{$ {\rm E_{B-V}}$}
\newcommand\teff{$ {\rm T_{eff}}$}

\newcommand\logg{$\log {\rm g}$}

\newcommand\loghe{$\log {\rm {\frac{N_{He}}{N_H}}}$}
\newcommand\roa{{\em ROA~5701}}
\newcommand\barn{{\em Barnard~29}}
\newcommand\bddre{{BD+33$^\circ$2642}}
\newcommand\klamma{\raisebox{1.5ex}[-2ex]{$\left] \rule{0pt}{3ex} \right.$}}
\newcommand\klammb{\raisebox{3ex}[-3ex]{$\left] \rule{0pt}{4.5ex} \right.$}}
\newcommand\klammc{\raisebox{4.3ex}[-4ex]{$\left] \rule{0pt}{6ex} \right.$}}
\begin{document}
\thesaurus{5(08.01.1 - 08.16.4 - 10.07.3 NGC 5139 - 10.07.3 NGC 6205)}
\title{Abundances of UV bright stars in globular clusters.
I. ROA 5701 in $\omega$ Centauri and Barnard~29 in M~13
\thanks{Based on observations collected a) with the NASA/ESA Hubble Space 
Telescope, obtained at the Space Telescope Science Institute, which is 
operated by the Association of Universities for Research in Astronomy, Inc., 
under NASA contract NAS 5-26555, and b) at the European 
Southern Observatory}}
\author{S. Moehler \and U. Heber \and M. Lemke \and R. Napiwotzki}
\offprints{S. Moehler}
\institute {Dr. Remeis-Sternwarte, Astronomisches Institut der Universit\"at 
Erlangen-N\"urnberg, Sternwartstr. 7, 96049 Bamberg, Germany}
\date{Received 18 June 1998/ accepted 27 August 1998}
\maketitle
\markboth{Moehler et al.: Abundances of UV bright stars in globular 
clusters. I.}
{Moehler et al.: Abundances of UV bright stars in globular clusters. I.}
\keywords{Stars: abundances -- Stars: AGB and post-AGB -- globular clusters 
individual: NGC 5139 -- NGC 6205}

\begin{abstract}

Two UV brights stars in globular clusters, \roa\ ($\omega$ Cen) and \barn\
(M~13) are analysed from high-resolution UV and optical spectra. The main  aim
is the measurement of iron abundances from UV spectra obtained with  the
HST-GHRS. In addition   atmospheric parameters and abundances for He, C, N, O,
and Si are  derived from optical spectra (ESO CASPEC) for \roa\ or taken
from  literature for \barn.   Both stars are found to be post-asymptotic giant
branch stars. Surprisingly, their iron abundances lie significantly below   the
cluster abundance in both cases. \barn\ lies 0.5 dex below the iron  abundance
derived for giant stars in M~13 and the iron abundance of \roa\  is the lowest
of any star in $\omega$ Cen analysed so far. \barn\ shows the same
abundance pattern as the red giant stars in M~13, except for its stronger iron
deficiency.  The iron depletion could be explained by gas-dust separation in
the AGB  progenitor's atmosphere, if iron condensed into dust grains which were 
then removed from  the atmosphere by a radiatively driven wind.  The
interpretation of the abundance pattern for \roa\ is hampered by the
star-to-star abundance variations seen in $\omega$ Cen, but its abundance 
pattern appears to be consistent with the gas-dust separation scenario. 
\end{abstract}

\section{Introduction}

The term ``UV bright stars'' was introduced by Zinn et al. (\cite{zine72})  for
stars in globular clusters that lie above the horizontal branch (HB) and are
bluer than red giants. The name resulted from the fact that in the U band these
stars were brighter than all other cluster stars. Further investigations showed
that this group of stars consists of post-AGB stars, evolving away from the
asymptotic giant branch (AGB) to the white dwarf domain (Sweigart et al.,
\cite{swme74}; de Boer \cite{debo87}), and of so-called Supra-HB stars that
evolve from the blue HB (BHB) towards higher luminosities (Dorman et al.,
\cite{doro93}) but do not have enough mass to ascend the AGB. Both evolutionary
stages are immediate precursors to white dwarfs. 

Up to now detailed analyses have been performed mainly for post-AGB stars in
the field of the Milky Way (McCausland et al., \cite{mcco92};  Napiwotzki et
al. \cite{nahe94}), for which the population membership is  difficult to
establish. The summarized result of these analyses is that the  abundances of
N, O, and Si are roughly 1/10 of the solar values, while Fe and  C are closer
to 1/100 solar. McCausland et al.\ (\cite{mcco92}) and Conlon (\cite{conl94})
interpret the  observed abundances as the results of dredge-up processes on the
AGB,  i.e. the mixing of processed material from the stellar interior to the
surface. Although this  hypothesis contradicts stellar evolution theories
(Renzini \& Voli \cite{revo81}; Iben \& Renzini \cite{ibre84}; Vassiliadis \&
Wood \cite{vawo93}), which do not predict any dredge-up processes for the
low-mass precursors of these objects,  compelling evidence that such dredge-up
processes do occur is provided by  K~648, the central star of the Planetary
Nebula Ps~1 in M~15.  Its atmosphere is strongly enriched in carbon when
compared to the cluster carbon  abundance (Heber et al., \cite{hedr93}) 
pinpointing the dredge-up of triple $\alpha$ processed material to the stellar
surface.

Napiwotzki et al.\ (\cite{nahe94}) discuss another possible explanation: The
photospheric abundances of the Pop~II central star BD+33$^\circ$2642 (and also
of the objects analysed by McCausland et al., \cite{mcco92}) 
can be understood as the results of gas-dust separation towards
the end of the AGB phase, which leads to a depletion of certain elements. This
process has already been proposed by Bond (\cite{bond91}) for some cooler
post-AGB stars with extreme metal deficiencies. A distinction between the two
scenarios is hampered by the fact that the original metallicities remain
unknown. Iron is one of the elements which are most sensitive to depletion.
Thus the knowledge of its abundance in the UV bright stars and a comparison
with the known cluster metallicity allows an unambiguous distinction between
both scenarios.  Unfortunately, for hot metal poor stars iron is not accessible
for a  spectroscopic analysis from optical spectra due to the lack of lines in
this wavelength region. In  the ultraviolet, however, a  large number of
spectral lines can be used. Hence the Goddard High  Resolution Spectrograph
(GHRS) of the Hubble Space Telescope was used to measure the iron line spectra 
of \roa\ and \barn .

\section{Observations and data reduction}

\subsection{UV spectroscopy} 

The UV spectra  of \roa\ and \barn\ were obtained with the Goddard High
Resolution Spectrograph onboard the Hubble Space Telescope,  equipped with the
G200M grating (1860 $-$ 1906 \AA,  0.07~\AA\ resolution)  and using the large
science aperture. This spectral region was chosen because the strongest
\ion{Fe}{III} absorption  lines are expected there as judged from the high
resolution IUE  spectra of \bddre\ (Napiwotzki et al., \cite{nahe94}),  which
shows an optical spectrum similar to \roa\ and \barn . The exposure times were
4711 s for \roa\ (observed on  August 3$^{\rm rd}$, 1996) and 4570 s for \barn\
(observed on November 30$^{\rm th}$, 1996). We did not use the FP split option
as we did not expect any lines strong  enough to allow a correct alignment of
the individual spectra by correlation.

After the standard pipeline reduction we co-added the flux of the individual
spectra, which were on identical wavelength scales. The resulting spectra
were converted to MIDAS bdf-format and interpolated to a step size of 0.02~\AA
. They were then corrected for Doppler shifts  using the heliocentric radial
velocities of the clusters (+232~km/sec, $\omega $ Cen; -246~km/sec, M~13) and
the corrections for heliocentric velocities  appropriate for the observation
dates. To allow a better definition of  the continuum we smoothed the spectra
with a 3 pixel wide box average filter.  The continuum was then defined by eye
and we estimate that the error of  the normalization lies between 5\% and 10\%.

Fig.~1 shows a section of the GHRS spectra of \roa\ and \barn\ compared to the
IUE data of \bddre\footnote{For this purpose the IUE  spectra SWP04791,
SWP37966, SWP43623,  SWP45533 were calibrated with the NEWSIPS software and
co-added}, a field  post-AGB star with \teff\ and \logg\ similar to \roa\ and
\barn .  The abundance noted for  \bddre\ in Fig.~1 was derived from
equivalent width measurements  (Napiwotzki et al., \cite{nahe94}) for a
microturbulent velocity of 10~km/s (which is the value Conlon et al.,
\cite{codu94}, derived for \barn .) The iron abundance of \bddre\ is  comparable
to that of M~13 and $\omega$ Cen. The much weaker iron lines in  the GHRS
spectra therefore suggest that \barn\ and \roa\ show significant iron 
depletions.

\begin{figure}
\vspace{6.8cm}
\includegraphics{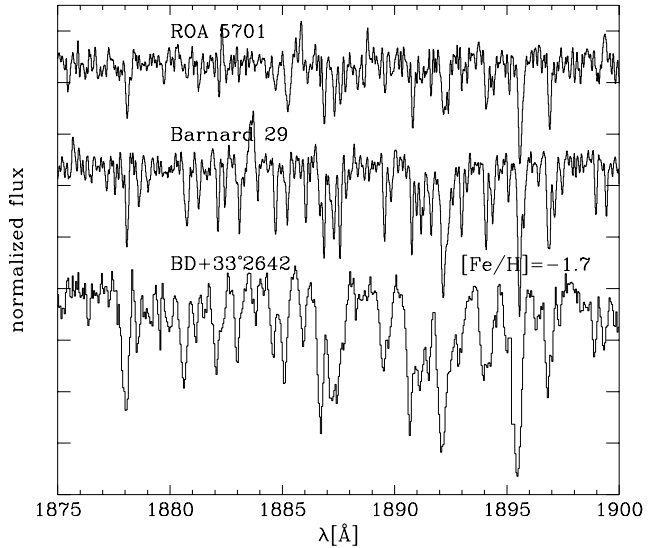}
\caption[]{The GHRS spectra of \roa\ and \barn , compared to IUE high 
resolution spectra of \bddre$^1$. The GHRS 
data were smoothed by convolution with a Gaussian of 0.06~\AA\ FWHM, which 
yields a ``smoothed'' resolution of 0.09~\AA , similar to that of the IUE 
data (0.1~\AA ). We used a microturbulent velocity of 10~km/s to 
derive the abundance noted for \bddre .}
\end{figure}

\subsection{Optical high resolution spectra}

\roa\ was observed with the ESO CAssegrain echelle SPECtrograph (CASPEC) at the
3.6m telescope at La Silla, Chile, on May 24, 1988. Two spectra of 1 hour
integration time each were obtained. The spectra were binned during  read out
in order to improve the S/N ratio. Reduction of the data proceeded in two
steps: first the ESO-MIDAS software  (Ponz \& Brinks, \cite{pobr86}) in
Garching was used for wavelength  calibration and extraction of the echelle
orders. The background correction  and flat fielding were done separately using
a computer program written by  G. Jonas (Kiel, see Heber et al.,
\cite{hewe88}). We then merged the orders of the CASPEC spectra and rebinned
them to a common wavelength scale. The spectra have a resolution of 0.3~\AA .

\subsection{Equivalent widths}

We always used the normalized spectra to measure equivalent widths.
The measurement in the optical spectra was 
straightforward since the lines are isolated and well defined and the  spectra
have a good S/N. In the UV,  however, the lines are more crowded and the S/N is
lower. Therefore,  three different methods were used: i) direct integration
without any fit of the line  shape using a global continuum (assuming that the
overall continuum  definition is more reliable than a local one due to the low
S/N),  ii) same as i), but for a local continuum, and iii) fitting Gaussians to
the absorption line profiles (using a locally  defined continuum).   Method
iii) could not be used for \roa\ because the lines were too weak.

For the GHRS data the equivalent widths measured for a global continuum (which
were used for the abundance determinations) were on average larger than those
measured for a  locally defined continuum. This offset leads to a difference in
the mean iron  abundance of 0.08 dex for \roa\ and 0.03~dex for \barn . 

For the CASPEC data the equivalent widths measured for a global continuum 
resulted in abundances larger than those determined for a local continuum  by
0.12~dex for O, 0.02~dex for N, and 0.14~dex for Si. As the CASPEC  spectra
showed small-scale continuum variations that are difficult to correct  for by a
global continuum fit we decided to keep the abundances derived from  the
``local continuum'' equivalent widths. 

\section{Effective temperatures, gravities and helium abundances}

Detailed analyses of optical spectra are available for \barn\ (Conlon et al.,
\cite{codu94}),  whereas only an estimate of the effective temperature from 
low resolution IUE spectra (Cacciari et al., \cite{caca84}, \teff\ = 24000~K,
\ebv\ $\ge$ \magpt{0}{11}) exists for \roa . Therefore we determined the
photospheric parameters  of \roa\ by analysing the optical CASPEC spectra
together with optical photometry (Norris \cite{norr74}) and low resolution IUE 
spectrophotometry (SWP07849, LWR06845). 

To derive a first estimate of the temperature we dereddened the optical 
photometry and the IUE data, using the interstellar extinction law of Savage \&
Mathis (\cite{sama79}). The UBV magnitudes were converted into fluxes, using
the conversion given by Heber et al. (\cite{hehu84}). The data were then fitted
with ATLAS9 (Kurucz \cite{kuru92}) models for [M/H] = $-$1.5. For a reddening
of \magpt{0}{11} (Norris, \cite{norr74}) resp. \magpt{0}{15} (Djorgovski,
\cite{djor93}) the best fits were achieved for effective temperatures of
22000~K resp. 24000~K. These results compare very well to those of Cacciari et
al. (\cite{caca84}).

For the temperatures given above we fitted ATLAS9 model spectra with solar
helium abundances and [M/H]~=~$-$1.5 to the Balmer lines  and obtained the
surface gravities from the best fit. 
To verify our results we also derived effective temperature, surface gravity,
and helium abundances from the Balmer and \ion{He}{I} line profiles only.  For
this purpose we calculated model atmospheres using ATLAS9 (Kurucz 1991, priv.
comm.) and used the LINFOR spectrum synthesis  program (developed originally by
Holweger, Steffen, and Steenbock at Kiel university) to compute a grid of
theoretical spectra, which include the Balmer lines H$_\alpha$ to H$_{22}$ and
\ion{He}{I} lines. The grid covered the range 10000 \ldots 27500~K in \teff,
2.5 \ldots 5.0 in \logg\ and $-$2.0 \ldots $-$0.3 in \loghe\ at a metallicity
of $-$1. For the actual fit we used the routines developed by Bergeron et al.
(\cite{besa92}) and Saffer et al. (\cite{saff94}), which employ a $\chi^2$
test. The results of all procedures are listed in Table~1. It is clear that the
agreement is rather good. For the further analysis we used the mean value of
the IUE derived results, i.e. an effective temperature of 23000~K and a log~g
value of 3.3.

\begin{table*}
\caption{Photospheric parameters of our programme stars.}
\begin{tabular}{l|lll|l} 
\hline
Name & \teff & \logg & \loghe & remarks \\
 & [K] & & & \\
\hline
ROA~5701 & 22000 & 3.2 & $-$1.00 & \ebv = \magpt{0}{11}, [M/H] = $-$1.5,
 fixed He abundance \\
ROA~5701 & 24000 & 3.4 & $-$1.00 & \ebv = \magpt{0}{15}, [M/H] = $-$1.5,
 fixed He abundance \\
ROA~5701 & 24500 & 3.4 & $-$0.98 & Balmer and \ion{He}{I} line profiles only, 
[M/H] = $-$1.0\\
\hline
Barnard 29 & 20000 & 3.0 & $-$1.06 & Conlon et al., \cite{codu94}, 
[M/H] = $-$1.0 \\
\hline
\end{tabular}
\end{table*}

\section{Abundance analysis}

The determination of elemental abundances is  interlocked with the
microturbulent velocity $\xi$, which  can be derived if a sufficient number of 
lines of one ion can be measured over a wide range of line strengths. In the 
optical \ion{O}{II} lines are most suitable for this purpose as they are  most
frequent. For \roa\ we could measure equivalent widths of  22 \ion{O}{II}
lines, which yield a value of 20~km/s for $\xi$.  Gies \& Lambert
(\cite{gila92}) note in their analysis of B-type  supergiants that the high
microturbulent velocities of about 20~km/s obtained from LTE analyses decrease
to about 10~km/s if NLTE effects are taken into account. The microturbulent
velocity of \barn,  $\xi$=10~km/s, has been determined by Conlon et al.
(\cite{codu94}). \footnote{ If we exclude the blend at $\lambda\lambda$ 1892.3
\AA\ from the line list the \ion{Fe}{III} lines yield 10~km/s for $\xi$ for
\barn . For \roa\ we derive a microturbulent velocity of 2--3~km/s from the 
measured \ion{Fe}{III} lines. Determinations of $\xi$ from model spectra to
which noise was added show that noise tends to increase the measured 
microturbulent velocity rather than decrease it.}

\subsection{Methods}

Abundances have been derived using the classical curve-of-growth technique  as
well as a spectrum synthesis technique. In both cases we computed  model
atmospheres for the appropriate values of effective temperature, surface
gravity, and cluster metallicity and used the LINFOR  spectrum synthesis
package for the further analysis.

\begin{description}
\item [curve of growth analysis]
We calculated curves of growth for the elements of interest, from which 
abundances were derived. We took into account that in many cases more than one 
line contributed to the measured equivalent widths by treating those lines as 
blends in LINFOR. In addition we tried to avoid lines with significant 
contributions from other elements or ions.

\item [spectrum synthesis]
In a second trial we fitted the whole spectrum at once. In this mode the 
LINFOR package tries to fit the line profiles of the metal lines using  a
$\chi^2$ test by adjusting the abundance of the element(s) that are fitted.  We
used the same line lists as for the curve-of-growth analysis. \end{description}

\begin{table}
\caption{Equivalent widths and abundances for \ion{C}{II} (upper limit only), 
\ion{N}{II}, \ion{O}{II}, and \ion{Si}{III} as derived from the CASPEC spectra 
of \roa .}
\begin{tabular}{rlrr|rr}
\hline
Ion/ & $\lambda$ & $\chi$ & log gf & W$_\lambda$ & log $\epsilon$ \\
Multiplet & [\AA] & [eV] & & [m\AA] &  \\
\hline
\ion{C}{II}
  6 & 4267.020        & 18.047  &  0.559 &        & \\
  6 & 4267.270\klamma & 18.047  &  0.734 & \hspace{2.2mm}$<$10  & $<$5.78  \\
\ion{N}{II}
 12 & 3995.000        & 18.498  &  0.225 & 38     & 6.79 \\
 15 & 4447.030$^1$    & 20.411  &  0.238 & 20     & 6.86 \\
\ion{O}{II}
 17 & 3919.285        & 25.655  & -0.247 & 36     & 8.29 \\
 10 & 4069.623        & 25.625  &  0.157 &        & \\
 10 & 4069.886\klamma & 25.632  &  0.365 & 86     & 7.98  \\
 10 & 4075.862        & 25.658  &  0.700 & 66     & 7.73 \\
 48 & 4089.285        & 28.699  &  0.885 & 59     & 8.22 \\
 20 & 4119.216        & 25.842  &  0.454 & 33     & 7.63 \\
 36 & 4189.600        & 28.354  & -0.821 &        & \\
 36 & 4189.789\klamma & 28.354  &  0.723 & 35     & 7.99 \\
 59 & 4302.858        & 31.311  &  0.092 &        & \\
 59 & 4303.070        & 31.311  & -0.008 &        & \\
  2 & 4303.836\klammb & 28.814  &  0.660 & 37     & 8.17 \\
  2 & 4345.559        & 22.973  & -0.330 & 36     & 7.80 \\
  2 & 4349.426        & 22.993  &  0.085 & 54     & 7.62 \\
  2 & 4366.888        & 22.993  & -0.319 & 38     & 7.82 \\
  5 & 4414.901        & 23.435  &  0.211 & 71     & 7.79 \\
  5 & 4416.973        & 23.413  & -0.041 & 54     & 7.86 \\
 32 & 4447.673$^1$    & 28.354  & -1.380 &        & \\
 32 & 4448.186$^1$\klamma & 28.354  &  0.047 & 20     & 7.79 \\
 14 & 4590.972        & 25.655  &  0.346 & 53     & 8.06 \\
 14 & 4595.960        & 25.655  & -1.037 &        & \\
 14 & 4596.176\klamma & 25.655  &  0.196 & 41     & 8.03 \\
  1 & 4638.857        & 22.960  & -0.307 & 41     & 7.90 \\
  1 & 4641.817        & 22.973  &  0.084 & 78     & 7.91 \\
  1 & 4649.143        & 22.993  &  0.343 & 112    & 7.93 \\
  1 & 4650.842        & 22.960  & -0.331 & 45     & 7.97 \\
  1 & 4661.633        & 22.973  & -0.249 & 61     & 8.08 \\
  1 & 4676.236        & 22.993  & -0.359 & 41     & 7.96 \\
 35 & 4699.003        & 28.502  &  0.429 &        & \\
 22 & 4699.220\klamma & 26.219  &  0.270 & 41     & 7.99 \\
 22 & 4705.350        & 26.242  &  0.518 & 43     & 7.93 \\
\ion{Si}{III}
  2 & 4552.620        & 19.018  &  0.283 & 64     & 6.00 \\
  2 & 4567.820        & 19.018  &  0.061 & 54     & 6.13 \\
  2 & 4574.760        & 19.018  & -0.509 & 17     & 6.13 \\
\hline
\end{tabular}\\
\begin{tabular}{l}
$^1$ blend of \ion{N}{II}/\ion{O}{II} (not used for the determination of 
$\xi$)\\
\end{tabular}
\end{table}

\subsection{Analysis of the optical spectrum of \roa}

For the analysis of the CASPEC data of \roa\  we used atomic data for
\ion{C}{II} (Yan et al.,  \cite{yata87}), \ion{N}{II} (Becker \& Butler,
\cite{bebu89}), \ion{O}{II}  (Bell et al., \cite{behi94}) and \ion{Si}{III}
(Becker \& Butler, \cite{bebu90}). Table~2 lists the results for individual
lines. For \ion{C}{II} we could  only derive an upper limit, assuming an
equivalent  width of 10~m\AA\ for the \ion{C}{II} line at 4267~\AA . The spectrum
synthesis resulted in abundances higher than those from the classical
curve-of-growth analysis by about 0.1~dex for \ion{N}{II}, \ion{O}{II}, and
\ion{Si}{III} (cf.  Table~5). Part of this offset is due to the fact that the
spectrum  synthesis has to use a ``global continuum'' for its fit
(see also Sect.~2.3).

\begin{table*}
\caption{Equivalent widths and abundances as derived from the GHRS spectra 
of \barn\ and \roa . We list the equivalent widths that were measured for a 
global continuum. Multiplet numbers are from Ekberg (1993).}
\begin{tabular}{rlrr|rr|rr}
\hline
Ion/ & $\lambda$ & $\chi$ & log gf & \multicolumn{2}{|c|}{ROA 5701} &
\multicolumn{2}{|c}{Barnard 29}\\
Multiplet & & & & W$_\lambda$ & log $\epsilon$ & W$_\lambda$ & log $\epsilon$ \\
 & [\AA] & [eV] & & [m\AA] &  & [m\AA] &  \\
\hline
\ion{Fe}{III} 
 UV52 & 1866.315 &         7.869 & -0.249 & 29 & 4.88 & & \\ 
 UV52 & 1866.570 &         7.870 & -0.770 & 25 & 5.33 &    & \\ 
 UV52 & 1869.837 &         7.870 & -2.134 &    &      &    & \\
      & 1869.844 &         7.870 & -2.805 &    &      &    & \\
      & 1869.927\klammb & 11.578 &  0.123 &    &      & 62 & 5.93 \\ 
      & 1870.592 &         8.245 & -1.665 &    &      &  6 & 5.58 \\
      & 1871.336        & 10.332 & -0.646 &    &      &    & \\ 
      & 1871.446\klamma & 10.332 & -0.862 & 17 & 5.47 &    &  \\ 
      & 1872.211 &        11.144 &  0.243 & 21 & 5.09 & 54 & 5.64 \\ 
      & 1873.549 &        10.317 & -0.737 &    &      & 14 & 5.64 \\
 UV62 & 1878.003 &         8.639 &  0.312 & 50 & 4.80 & 54 & 4.87 \\ 
      & 1878.547 &         8.238 & -0.991 &    &      & 27 & 5.63 \\
      & 1881.194 &        11.118 &  0.225 & 15 & 4.94 & 30 & 5.27 \\ 
 UV62 & 1882.050 &         8.639 & -0.082 &    &      &    & \\
      & 1882.064\klamma &  8.650 & -1.339 &    &      & 39 & 5.03 \\ 
 UV62 & 1884.597 &         8.656 & -0.257 & 26 & 5.05 & 60 & 5.51 \\ 
 UV96 & 1885.965 &        10.308 & -0.083 & 12 & 4.94 & 34 & 5.45 \\ 
      & 1886.609 &         8.656 & -0.316 &    &      &    &  \\ 
 UV52 & 1886.765 &         7.867 &  0.336 &    &      &    & \\ 
      & 1886.783\klammb & 11.144 & -1.870 & 34 & 4.31 & 81 & 4.78 \\ 
 UV53 & 1887.212 &         7.867 & -0.068 & 37 & 4.82 & 69 & 5.21 \\ 
      & 1887.472 &         7.870 & -0.464 &    &      &    & \\ 
 UV52 & 1887.479 &         7.870 & -1.541 &    &      &    & \\ 
      & 1887.492\klammb &  7.869 & -0.286 & 25 & 4.61 & 53 & 4.99 \\ 
      & 1887.751 &        14.170 &  0.484 &    &      &    & \\ 
      & 1887.752\klamma &  7.870 & -1.204 & 13 & 5.11 & 16 & 5.19 \\ 
 UV53 & 1889.463 &         7.870 & -0.302 &    &      &    & \\ 
      & 1889.471\klamma &  7.870 & -0.889 & 12 & 4.44 & 37 & 4.92 \\ 
      & 1889.742 &        10.765 & -0.141 &    &      & 22 & 5.39 \\
 UV52 & 1890.678 &         7.861 &  0.504 & 40 & 4.29 &    & \\
 UV53 & 1890.882 &         7.870 & -0.805 &    &      &    & \\
      & 1890.882\klamma & 10.368 & -2.528 &    &      & 28 & 5.36 \\ 
 UV52 & 1892.151 &         7.867 & -0.355 &    &      &   & \\
 UV96 & 1892.253 &        10.308 &  0.140 &    &      &   & \\
      & 1892.348\klammb & 10.994 & -0.409 &    &      & 197 & 6.08 \\
      & 1892.882 &         7.869 & -1.205 &    &      &    & \\ 
 UV96 & 1892.896 &        10.308 & -0.047 &    &      &    & \\ 
      & 1893.111\klammb & 13.127 &  0.116 & 28 & 5.10 & 42 & 5.27 \\ 
      & 1893.309 &        10.368 & -0.387 &    &      & 12 & 5.21 \\
 UV83 & 1893.988 &         9.899 &  0.481 & 36 & 4.80 & 46 & 4.95 \\ 
      & 1896.334 &        14.174 &  0.279 &    &      & 13 & 5.63 \\
      & 1896.741 &        10.225 & -0.059 &    &      &    & \\ 
 UV83 & 1896.821\klamma &  9.897 &  0.514 & 51 & 4.84 & 105 & 5.44 \\ 
      & 1897.385 &         9.899 & -0.677 &    &      & 20  & 5.63 \\
      & 1901.258 &        10.305 & -0.691 &    &      &    & \\ 
      & 1901.383 &        10.225 & -0.517 &    &      &    & \\ 
      & 1901.394 &        10.211 & -0.073 &    &      &    & \\ 
 UV96 & 1901.549\klammc & 10.368 & -1.813 & 31 & 5.14 & 95 & 5.83 \\ 
 UV94 & 1902.098 &        10.308 &  0.118 &    &      & 36 & 5.28 \\
      & 1902.411 &        10.215 & -0.051 &    &      & 44 & 5.54 \\
      & 1902.910 &        10.895 &  0.243 &    &      & 30 & 5.20 \\
      & 1903.177 &        10.308 & -1.003 &    &      &    & \\
      & 1903.263\klamma &  9.153 & -0.547 &    &      & 49 & 5.70 \\
      & 1904.265 &        13.581 &  0.224 &    &      &    & \\
      & 1904.384 &        13.127 & -0.235 &    &      &    & \\
      & 1904.412\klammb &  8.656 & -0.274 &    &      & 58 & 5.37 \\ 
\hline
\end{tabular}
\end{table*}

\subsection{Iron abundances}

Our main aim is to determine the iron abundances of both stars from the 
\ion{Fe}{III} lines in the UV. We used the atomic data given by Ekberg 
(\cite{ekbe93}). The measured equivalent widths and the resulting 
abundances for both stars are listed in Table~3.
The spectrum synthesis yields iron abundances  about 
0.2~dex lower than those obtained from the equivalent widths. 

\begin{table*}
\caption{Error estimates}
\begin{tabular}{lr|rrr|rrr}
\hline
Star & Ion & \multicolumn{3}{|c}{$\Delta\log\epsilon$ from W$_\lambda$} & 
 \multicolumn{3}{|c}{$\Delta\log\epsilon$ from spectrum synthesis}\\
 &  & $\pm$1000~K & $\pm$0.1 dex & $\pm$5~km/s & $\pm$1000~K & $\pm$0.1 dex 
 & $\pm$5~km/s\\
\hline
ROA 5701 & \ion{C}{II}   & $\pm$0.08 & $\mp$0.03 & $\mp$0.01
                         & $\pm$0.03 & $\mp$0.03 & $\pm$0.05\\
         & \ion{N}{II}   & $\pm$0.00 & $\mp$0.00 & $\mp$0.01
                         & $\pm$0.02 & $\mp$0.01 & $\mp$0.01\\
         & \ion{O}{II}   & $\mp$0.12 & $\pm$0.04 & $\mp$0.04
                         & $\mp$0.13 & $\pm$0.01 & $\mp$0.02 \\
         & \ion{Si}{III} & $\mp$0.06 & $\pm$0.03 & $\mp$0.02
                         & $\mp$0.07 & $\pm$0.03 & $\pm$0.01 \\
         & \ion{Fe}{III} & $\pm$0.03 & $\pm$0.02 & $\mp$0.01
                         & $\mp$0.04 & $\pm$0.04 & $\pm$0.02 \\
\hline
Barnard 29 & \ion{Fe}{III} & $\mp$0.04  & $\pm$0.04 & $\mp$0.11
                           & $\mp$0.05  & $\pm$0.04 & $\mp$0.08 \\
\hline
\end{tabular}
\end{table*}

\subsection{Error estimates}

The iron abundances derived for \roa\ and \barn\ from different equivalent 
width measurements  differ by up to 0.1~dex and 0.05~dex, respectively.  The
effects of differences in effective temperature, surface gravity, and 
microturbulent velocity are given in Table~4. To check the effects of using
different line lists we also derived  abundances using the line lists of Kurucz
(1991, priv. comm., observed  lines only). The derived mean abundances differed
by about 0.05~dex from those  given in Table~5. The errors given in Table~5
include those of Table~4 plus 0.05~dex (to account for possible errors in the
line lists) and 0.1~dex resp. 0.05~dex in the iron abundances derived from
W$_\lambda$ (to account for errors in the equivalent widths measurements).

The stars we analyse are in a temperature-gravity range where NLTE  effects
start to play a r\^ole. Gies \& Lambert (\cite{gila92}) and Kilian 
(\cite{kili94}) find that the NLTE abundances of C, N, O, and Si for 
B-type main sequence and supergiant stars are on average about 0.1 -- 0.2~dex 
lower than the corresponding LTE abundances. As the NLTE corrections shift the 
abundances of all elements in the same direction the relative abundance trends 
remain unchanged within our error bars.

Prompted by a remark from the referee Dr. P. Dufton we  investigated
whether the low iron abundances that we derived from UV lines of \ion{Fe}{III} 
might be due to systematic errors. For this check we obtained 6 high 
resolution  UV spectra  of the normal main sequence B star  $\gamma$~Peg  from
the IUE final archive, which we coadded in order to increase the S/N.  
As line blending, continuum placement and incompleteness of atomic line lists 
are much more severe problems for solar metallicity B stars than 
for our (metal-poor) programme stars, we synthesized the spectral range in 
question  using the
entire Kurucz line list and varied the iron abundance. From the  IUE data we
get an iron  abundance of $\log{\epsilon_{Fe}} = 6.95$\footnote{An error of 5\%
in the  continuum definition results in an error of 0.2~dex in the iron
abundance.}.  This is at variance with the near solar value
($\log{\epsilon_{Fe}} = 7.56$, Pintado \& Adelman, \cite{piad93}) derived from
the analysis of optical \ion{Fe}{III} lines for this star, which we confirm
from ESO CASPEC spectra. A similar discrepancy has been found by Grigsby et
al. (\cite{grmu96})  for the normal main sequence B star $\iota$ Her, for which
the iron abundance from  ultraviolet \ion{Fe}{II} and \ion{Fe}{III}
lines was found to be more  than 0.47 dex below the near solar value  obtained
from the analysis of optical lines.  Grigsby et al. argue that the lower iron 
abundance derived from the UV lines is the correct one, as the 
abundances for both ions agree,  whereas the previous optical analyses found
differences between the two  ions of up to 1.0 dex. 

In summary the UV spectral analysis of normal B stars is currently so
severely hampered by line crowding, continuum definition and incompleteness of 
atomic data that it is impossible to draw any clear cut conclusions on
systematic  abundance offsets for our metal poor programme stars, which are 
much less affected by these problems. We therefore
do not apply any offsets to the iron abundances derived from the GHRS spectra.

\begin{table*}
\caption{Abundances derived for \roa\ and \barn . $\log{\epsilon}$ gives the 
number abundance of the respective element with $\log{\epsilon}$ = 
$\log{(X/H)}+12$.}
\begin{tabular}{rrrrr|rrrrr|l}
\hline
 \multicolumn{5}{c}{ROA 5701} & \multicolumn{5}{|c|}{Barnard 29 } & remarks\\
\hline
 $\log{\epsilon_C}$ & $\log{\epsilon_N}$ & $\log{\epsilon_O}$ & 
 $\log{\epsilon_{Si}}$ & $\log{\epsilon_{Fe}}$ & $\log{\epsilon_C}$ & 
 $\log{\epsilon_N}$ & $\log{\epsilon_O}$ & $\log{\epsilon_{Si}}$ & 
 $\log{\epsilon_{Fe}}$ & \\
\hline
 $<$5.78   & 6.83          & 7.93          & 6.08          & 4.89 & 
           & 7.30$^1$      & 7.60$^1$      & 6.30$^1$      & 5.38 & 
 W$_\lambda$\\
 $\pm$0.10 & $\pm$0.05     & $\pm$0.14     & $\pm$0.09     & $\pm$0.12 & 
 & $\pm$0.11$^1$ & $\pm$0.20$^1$ & $\pm$0.27$^1$ & $\pm$0.14 & \\
 & & & & & & & & & & \\
 $\le$5.92 & 6.89          & 8.00          & 6.20          & 4.68      &
 6.15$^2$  & & & & 5.21 & spectrum synthesis\\
 $\pm$0.08     & $\pm$0.06 &  $\pm$0.14  &  $\pm$0.09    & $\pm$0.08 &
 $\pm$0.10$^2$ &           &             &  & $\pm$0.10    & \\
\hline
 8.58 & 8.05 & 8.93 & 7.55 & 7.50 & 8.58$^{\ }$ & 8.05$^{\ }$ & 8.93$^{\ }$ & 
 7.55$^{\ }$ & 7.50 & solar values \\
\hline
 [C/H]    & [N/H]   & [O/H]   & [Si/H]  & [Fe/H]  & 
 [C/H]    & [N/H]   & [O/H]   & [Si/H]  & [Fe/H]  & \\
\hline
 $<-$2.80 & $-$1.22 & $-$1.00 & $-$1.47 & $-$2.61 & & $-$0.75$^1$ & 
 $-$1.33$^1$ & $-$1.25$^1$ & $-$2.12 & W$_\lambda$\\
  $\le-$2.66 & $-$1.16 & $-$0.93 & $-$1.35 & $-$2.82 & $-$2.33$^2$ & & & & 
  $-$2.29 & spectrum synthesis\\
\hline
\end{tabular}\\
$^1$ Conlon et al. (1994)\\
$^2$ Dixon \& Hurwitz (1998)\\
\end{table*}

\begin{figure} 
\vspace{6.5cm} 
\includegraphics{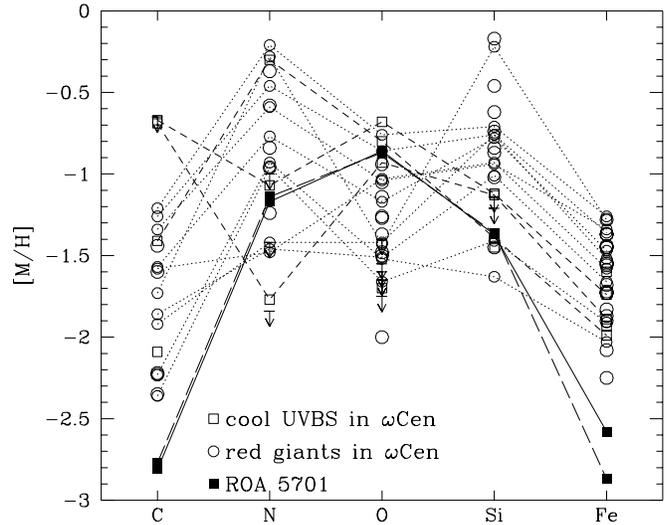} 
\caption[]{The abundances derived for \barn\  compared to those of red giant
stars in M~13. The solid line connects the abundances derived for \barn\  from
curve-of-growth analyses. The dotted and short-dashed  lines mark the
abundances of the red giants taken from Smith et al. (1996) resp. Kraft et al.
(1997).} 
\end{figure}

\begin{figure}
\vspace{6.5cm}
\includegraphics{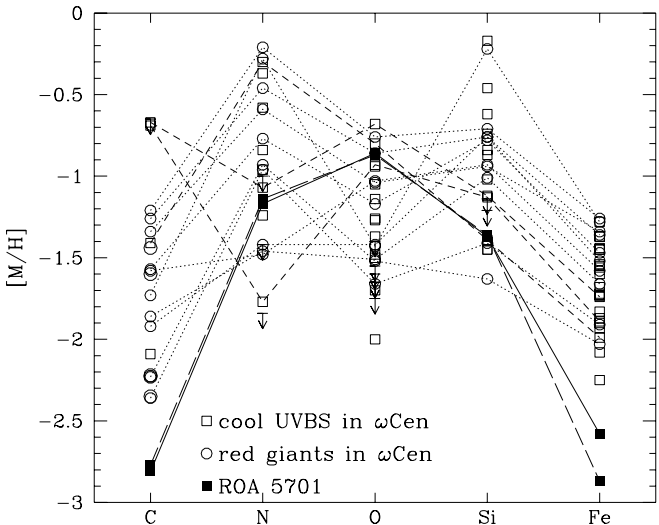}

\caption[]{The abundances derived for \roa\  compared to those of other stars
in $\omega$ Cen. The solid and long-dashed lines connect the abundances 
derived for \roa\  from curve-of-growth and spectrum synthesis analysis,
respectively. The  short dashed  lines mark the cool UV bright stars, the
dotted lines connect  the  abundances of the red giants. For references see
text. } \end{figure}

\section{Discussion}

The results of our abundance analyses are given in Table~5, where we also
give the results of Conlon et al. (\cite{codu94}) and Dixon \& Hurwitz
(\cite{dihu98}) for \barn . A comparison of our results with RGB and cooler UV
bright  stars (only in $\omega$ Cen) is presented in Figs. 2 and 3. The M~13
data in Fig.~2 are from Smith et al.  (\cite{smsh96}) and Kraft et al.
(\cite{krsn97}). The $\omega$ Cen data for the giants in Fig.~3 were collected
from Paltoglou \& Norris (\cite{pano89}), Brown et al. (\cite{brwa91}), Brown
\& Wallerstein (\cite{brwa93}), and Smith et al. (\cite{smcu95}), the data for
the cool UVBS are from Gonzalez \& Wallerstein (\cite{gowa94}).   All
abundances were adjusted to a solar iron abundance of
$\log{\epsilon_{Fe}}$~=~7.50.

As can be seen from Fig.~2 \barn\ shows similar abundance trends (except for 
iron) as the giant  stars observed in M~13. This abundance pattern is likely
caused by deep  mixing and dredge-up of CNO-processed material on the first RGB
(Pilachowski  et al., \cite{pisn96}). However, the iron abundance we found for
\barn\ is lower than the cluster value by 0.5\,dex. This indicates that the
atmosphere of \barn\ has become iron depleted during the star's evolution, most
likely by the  gas-dust separation proposed by Bond (\cite{bond91}) and 
discussed by Napiwotzki et al.\ (\cite{nahe94}), which is also favoured by
Dixon \& Hurwitz (\cite{dihu98}) for \barn . This scenario assumes that the
dust formed in the cool and extended atmospheres  of AGB stars is selectively
removed by radiation pressure, while the gas remains at the stellar surface.
Since metals with high condensation temperatures (e.g. iron) preferentially
condense into dust grains and elements with lower condensation temperatures
remain in the gas phase, the remaining gas forms  a iron-poor atmosphere.
Mathis \& Lamers (\cite{mala92}) discussed a single-star and a binary
scenario, which can both result in a selective removal of dust at the end of the
AGB stage.

\begin{figure}
\vspace{10.5cm}
\includegraphics{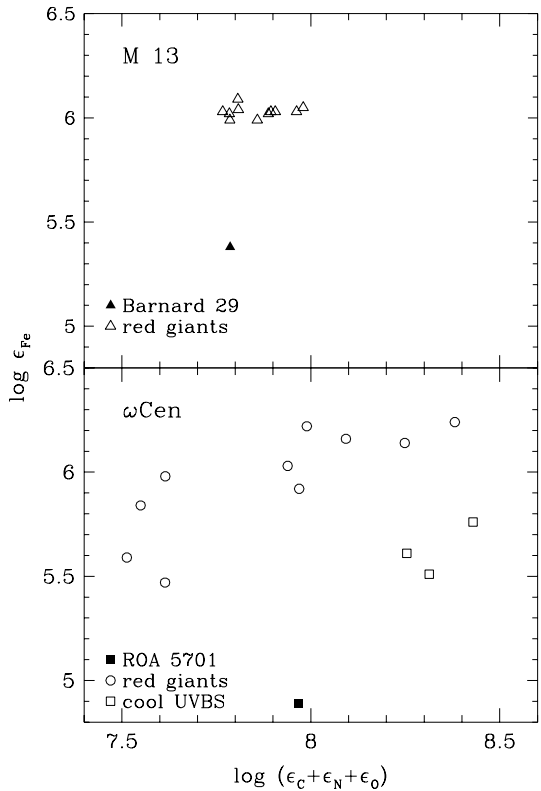}
\caption[]{
{\bf Upper panel:} The Fe abundances vs. the sum of CNO abundances for 
\barn\ compared to those of red giants in M~13. For references see text.
{\bf Lower panel:} The Fe abundances vs. the sum of CNO abundances 
for \roa\ compared to those of giants and cool UVBS in $\omega$ Cen.
For references see text.}
\end{figure}

The interpretation of the abundance pattern of \roa\ is hampered by the complex 
patterns found in $\omega$\,Cen stars in general (cf.\ Fig.~3). If we take as a
metallicity tracer the sum of  C\footnote{One should keep in mind that the 
upper limit for the C abundance was derived solely  from the 4267~\AA\ line,
for which  NLTE effects (Eber \& Butler, \cite{ebbu88}) can lead to  abundances
too low by up to 0.5~dex.},  N, and O abundances (which remains unchanged  by
CNO processing), we find that the original metallicity of \roa\ is close
to the median value determined for cluster stars (cf. Fig.~4). However, the
iron abundance  is 0.5~dex below the lowest value found for any RGB star
plotted in Fig.~3.  This points to an iron depletion similar to that detected in
\barn , which also shows a similar behaviour in Fig.~4.

In summary, the C, N, O, and Si abundances of \barn\ and \roa\ are in agreement
with that of the respective cluster red giant stars. No dredge-up during the
AGB phase is necessary to explain these abundance patterns, although a moderate
third dredge-up cannot be ruled out  from our data alone. However, our low iron
abundances point towards a significant iron depletion, most probably  caused by
a gas-dust separation during the late AGB stages.

\acknowledgements 
SM acknowledges financial support from the DARA under grant 50~OR~96029-ZA.
We want to thank the referee Dr. P. Dufton for his valuable advice.

\end{document}